# Narrowband, angle-stable, and highly efficient polariton organic light emitting diodes employing thermally activated delayed fluorescence

Andreas Mischok[1,#,*], Simon Lennartz[1,#], Vanessa Gruber[1], Francisco Tenopala-Carmona[1], Julia Witt[1], Sabina Hillebrandt[1], Malte C. Gather[1,2,*]

[1]*Humboldt Centre for Nano- and Biophotonics, Department of Chemistry, University of Cologne, Greinstr. 4-6, 50939 Köln, Germany*

[2]*School of Physics and Astronomy, University of St Andrews, North Haugh, St Andrews KY16 9SS, United Kingdom*

[#]*These authors contributed equally.*

*contact: andreas.mischok@uni-koeln.de, malte.gather@uni-koeln.de

## Abstract

Narrowband emission is crucial for next generation optoelectronic devices to satisfy demands for high color brilliance. Microcavities can narrow emission spectra of organic light-emitting diodes (OLEDs) through the creation of resonant standing waves, independent of emitter material, enabling flexibility in molecular and device design. However, they induce a strong angle-dependence of the perceived emission color. Utilizing strong light-matter coupling of cavity photons with the virtually angle-independent exciton, leads to exciton-polariton emission showing reduced linewidths and suppressed angular dispersion if tuned correctly. Creating polaritons in highly efficient materials such as thermally activated delayed fluorescence (TADF) molecules is however difficult as their low oscillator strengths intrinsically disfavor light-matter interaction. Here, we present the successful combination of a highly efficient but intrinsically broadband TADF emitter with a strongly absorbing assistant strong coupling material in a modified microcavity structure. Through optimizing the assistant strong coupling layer architecture, we demonstrate polariton OLEDs with exceptionally narrowband, angle stable emission at external quantum efficiencies above 20% for both bottom- and top-emitting designs, more than doubling the performance of previous record devices. The results pave the way for utilizing polaritonic emission at practical device efficiencies for display applications.

## Introduction

Organic semiconductors exhibit a multitude of beneficial properties for both scientific research and every-day applications. In particular, they are able to sustain stable excitons at room temperature due to their large exciton binding energies, and they offer high oscillator strength. Together, this facilitates light-matter interaction and the realization of high impact applications such as organic lasers[1,2], organic solar cells[3] and organic light-emitting diodes[4] (OLEDs), among others. OLEDs have now become essential for displays in high-end consumer devices, owing to their versatility, high efficiency and easy fabrication. A major breakthrough in OLED technology has been the utilization of triplet excitons. Excitons are formed through the combination of electrons and holes, which are injected from respective transport layers, in the active emissive layer. Due to spin statistics, this leads to the formation of singlet and triplet excitons in a 1:3 ratio. Consequently, OLEDs harvesting only singlet excitons are generally limited to an internal quantum efficiency of 25%. In order to harvest triplets, two major strategies have been established – enabling phosphorescence from the excited triplet to the singlet ground state through spin-orbit coupling in heavy metal complexes[5], or, more recently, enhancing reverse intersystem crossing (RISC) from the excited triplet back to the excited singlet state, leading to delayed fluorescence[6,7]. To enable efficient RISC, the energy gap between singlet and triplet excitons ($\Delta E_{ST}$) should be minimized, which is most often achieved by designing molecules with a spatially separated highest occupied molecular orbital (HOMO) and lowest unoccupied molecular orbital (LUMO). This so-called donor-acceptor design[8] enables $\Delta E_{ST}$ in the range of 10s of meV, and unlike for phosphorescence, does not rely on the use of scarce heavy metals. Consequently, RISC can be thermally activated at room temperature ($k_B T_{RT} \approx 25$ meV). This thermally activated delayed fluorescence (TADF) can achieve near 100% internal quantum efficiency in OLEDs and has become a major focus of research and development given its large possible application space. The donor-acceptor design of TADF molecules, however, means their energy levels are highly dependent on disorder in molecular conformation and local dielectric environment which, in addition to the presence of vibronic transitions, leads to large inhomogeneous broadening of emission[9,10], routinely showing spectral widths in the range of 50-100 nm. This is in stark contrast to the current demands for ultra-pure emission spectra in the display industry. Further, the limited overlap of HOMO and LUMO decreases the oscillator strength of TADF emitters and therefore the light-matter interaction in these materials.

Strong light-matter interaction has become an attractive playground for photonic systems. When the interaction rate between a photon in a high-quality microcavity (MC) and an exciton resonance becomes stronger than their individual loss rates, they form exciton-polariton (or simply polariton) quasiparticles[11]. Akin to molecular orbitals, strong coupling leads to the hybridization of energy levels, most commonly into a lower and upper polariton branch (LPB and UPB), separated by the Rabi splitting energy ($\hbar\Omega_R$). Because of this coherent light-matter interaction, polaritonic systems have shown great versatility for the creation of ultra-low-threshold polariton lasers[2,12], quantum simulators[13] and quantum information platforms[14], polariton switches as well as in polariton chemistry[15,16]. Here, organic semiconductors offer great advantages especially for practical devices that generally require operation at room temperature, which has led to the exploration of applications such as polaritonic solar cells[17], photodetectors[18,19], polariton transistors[20] and polariton OLEDs[21–23]. Polariton OLEDs (P-OLEDs) are attractive as they provide an easily controllable on-demand source of polaritons, and also because they exhibit narrowband tunable emission with ultralow angular dispersion, i.e., with a very weak change in emission spectrum across different observation angles. The latter is particularly attractive for next generation information displays with highly saturated color rendering but requires the design of highly efficient P-OLEDs, which has been a challenge in the past. While TADF is one of the most efficient OLED technologies, the low oscillator strength of TADF molecules seems to make them intrinsically unfavourable for polaritonics, however.

In this work, we demonstrate the design of highly efficient P-OLEDs based on a green TADF emitter through the use of an assistant strong coupling layer (SCL) with high oscillator strength. In order to achieve this combination, we expand an efficient TADF OLED stack by an electrically doped hole transport layer to enable the formation of a second order MC, adding the SCL in the second field maximum of the cavity to maximize light-matter interaction. The resulting TADF P-OLEDs show efficient emission from the lower polariton branch reaching a record external quantum efficiency (EQE) of over 20% at a luminance of 1,000 cd m$^{-2}$ in optimized designs, more than doubling the performance of previous P-OLEDs. We demonstrate strategies for flexible alteration of the SCL architecture, showcasing its facile implementation and how this will enable the use of SCLs in a wide variety of TADF systems. Finally, to provide evidence of the enormous potential TADF P-OLEDs offer for display applications, we demonstrate an optimized, narrowband and highly efficient top-emitting P-OLED with low angular dispersion.

**Results**

**Realizing TADF-Polariton-OLEDs**

Our P-OLED development is based on a TADF reference OLED comprising the highly efficient green TADF emitter 9,10-bis(4-(9H-carbazol-9-yl)-2,6-dimethylphenyl)-9,10-diboraanthracene (CzDBA), which exhibits low Δ$E_{ST}$, fast reverse intersystem crossing, near-unity photoluminescence quantum efficiency (PLQY), and favourable highly horizontal orientation (compare Wu *et al.*[24]). CzDBA has been used successfully in extremely efficient OLED designs, albeit with a rather broad emission linewidth on the order of 80 nm.

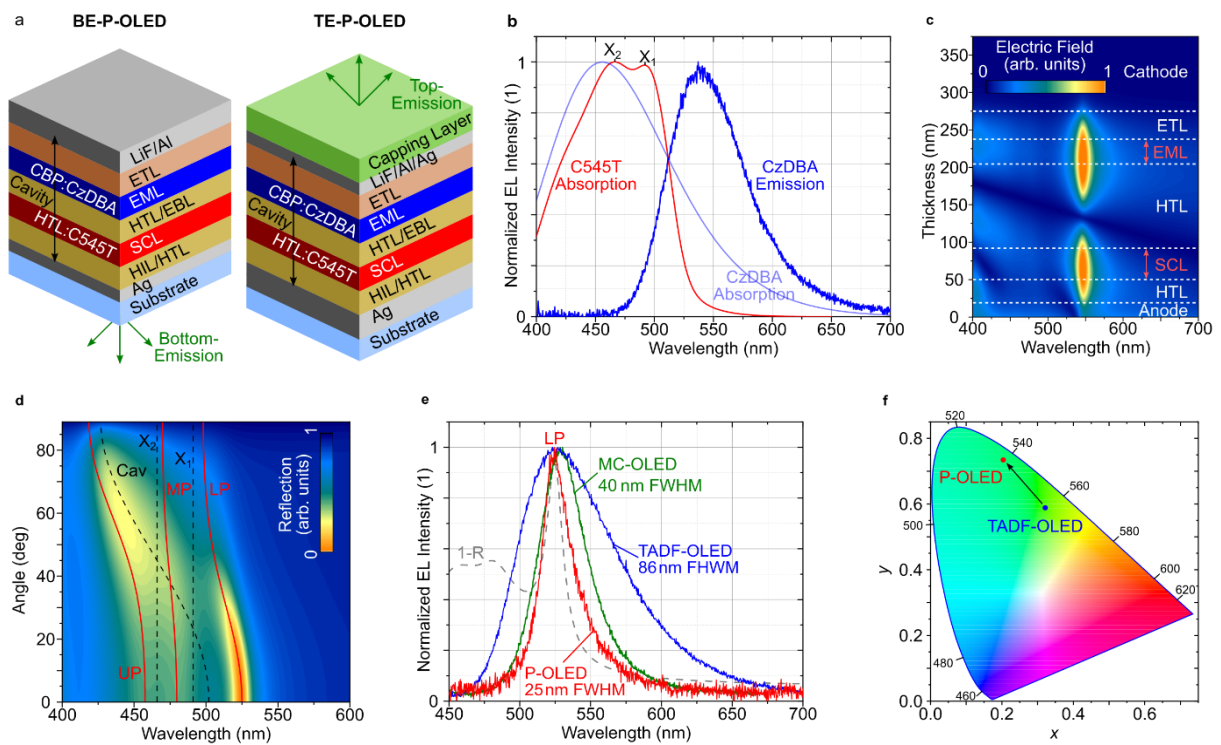

**Figure 1. Realization of TADF-Polariton-OLEDs. a** Layer stack of bottom- and top-emitting TADF P-OLED designs, combining the reference TADF-OLED and the assistant strong coupling layer (SCL) in a common cavity. Through the use of the SCL, polaritons are formed without a detrimental effect on

TADF in the emission layer. **b** Normalized absorption spectra of CzDBA (light blue) and C545T, and luminescence spectrum of CzDBA (dark blue). The small spectral shift between C545T absorption and CzDBA emission renders the assistant SCL effective. **c** Simulation of the electric field in a second order MC structure, showing field maxima at the positions of the EML and SCL, respectively. **d** Simulated reflection spectrum of a TADF P-OLED. Through coupling of the cavity photon with C545T excitons (dashed lines), a lower, middle and upper polariton branch are formed (red lines). **e** Comparison of electroluminescence spectra of a reference TADF-OLED with ITO anode (blue), a MC OLED with Ag anode (green) and a polariton OLED with Ag anode and SCL. The formation of polaritons enables a strong reduction in emission linewidth from 86 nm (full width at half maximum, FWHM) in the reference to 25 nm in the TADF P-OLED. Dashed line shows the inverted reflection spectrum of the TADF P-OLED, showing that its emission originates from the LP mode. **f** CIE 1931 coordinates for the reference TADF-OLED and the TADF P-OLED.

Figure 1b shows the intrinsic emission spectrum of CzDBA doped at 10wt% into a CBP host (full material names in Methods), showing green emission peaking at 540 nm, as well as the normalized absorption spectrum. The absorption of the intramolecular charge-transfer state between diboroanthracene (acceptor) and carbazole (donor) moieties of CzDBA, which peaks at 450 nm, is however very weak, showing a peak optical extinction coefficient (imaginary part of the complex index of refraction) of 0.009 in the doped film, which is approximately 100-fold lower than that of strong organic absorbers, such as the fluorescent emitter 2,3,6,7-tetrahydro-1,1,7,7,-tetramethyl-1H,5H,11H-10-(2-benzothiazolyl)quinolizino-[9,9a,1gh]-coumarin (C545T).

First, we fabricate a reference TADF-OLED with the following stack architecture: ITO (anode) | 50 nm NPB (HTL) | 10 nm TCTA (EBL) | 30 nm CBP:10wt% CzDBA | 40 nm TmPyPB (ETL) | 1 nm LiF (EIL) | 100 nm Al (cathode), where HTL - hole transport layer, EBL – electron blocking layer, ETL – electron transport layer and EIL – electron injection layer. This OLED shows green emission with a linewidth of 86 nm (Fig. 1e). In order to facilitate polariton formation, we modify the reference structure as follows: 1) creating a strong MC by replacing the ITO anode with a reflective and semi-transparent 1 nm Al | 20 nm Ag | 1 nm $MoO_3$ contact, 2) adding a highly conductive and relatively thick HTL comprising SpiroTTB, electrically doped with 4wt% F6TCNNQ as optical spacer to form a second order MC, and 3) introducing an SCL of C545T into the second electric field maximum of the mode supported by the MC. The resulting P-OLED device architecture is schematically shown in Fig. 1a for both the bottom- and top-emitting configuration investigated in this study. We utilize the second order MC to spatially separate the emission layer and SCL, while allowing for both of them to be placed in an electric field maximum of the cavity mode at the dominant emission wavelength of CzDBA (see calculations for bottom-emitting cavity in Fig. 1c). This maximizes both outcoupling of photons generated in the emission layer as well as light-matter interaction of photons with C545T excitons to generate polaritons. Transfer-matrix simulations of the reflection spectrum for the bottom-emitting P-OLED device stack show the telltale signs of polariton formation (Fig. 1d), namely a hybridization of the original cavity resonance (Cav) and the singlet 0-0 and 0-1 exciton transitions of C545T ($X_1$, $X_2$) into three polariton modes. The lower polariton (LP), middle polariton (MP), and upper polariton (UP), are also modelled using a three-coupled oscillator Hamiltonian between Cav, $X_1$ and $X_2$, and the results of this coincides with the position of the minima in the reflection spectrum calculated by transfer-matrix.

Comparing electroluminescence spectra of the reference TADF-OLED with an ITO anode to the MC cavity (with Ag anode but without SCL) and the P-OLED (with Ag anode and SCL), we observe a drastic reduction in the full width at half maximum (FWHM) linewidth of the emission, initially from the MC resonance (from 86 nm to 40 nm) and then even further by polariton formation (to 25 nm). This additional reduction in linewidth when going from a weakly coupled MC to a polariton mode has been observed in organic cavities[25,26] before and was interpreted as a reduction of the lower polariton decay rate through exciton-photon interaction[27,28]. The drastic reduction in linewidth increases color purity and in turn shifts the Commission internationale de l'éclairage (CIE) 1931 coordinates towards pure green emission (Fig. 1f).

**Microcavity vs Polariton-OLEDs**

Figure 2 shows the transition from MC OLED to P-OLED in more detail. Relative to the literature known reference TADF device (compare Wu *et al.*[24]), we increase the overall device thickness by inserting a 140 nm p-doped HTL of SpiroTTB:4wt% F6TCNNQ. As the conductivity of the doped film is several orders of magnitude higher than the undoped layers, its inclusion only minimally affects hole transport to the EML of the OLED. We can therefore increase the overall cavity thickness to the second order, without negatively impacting the charge carrier balance in the EML. The microcavity OLED shows a narrowed emission with a peak at 550 nm, and the typical parabolic angular dispersion of a MC (Fig. 2a), both in reflection and in electroluminescence.

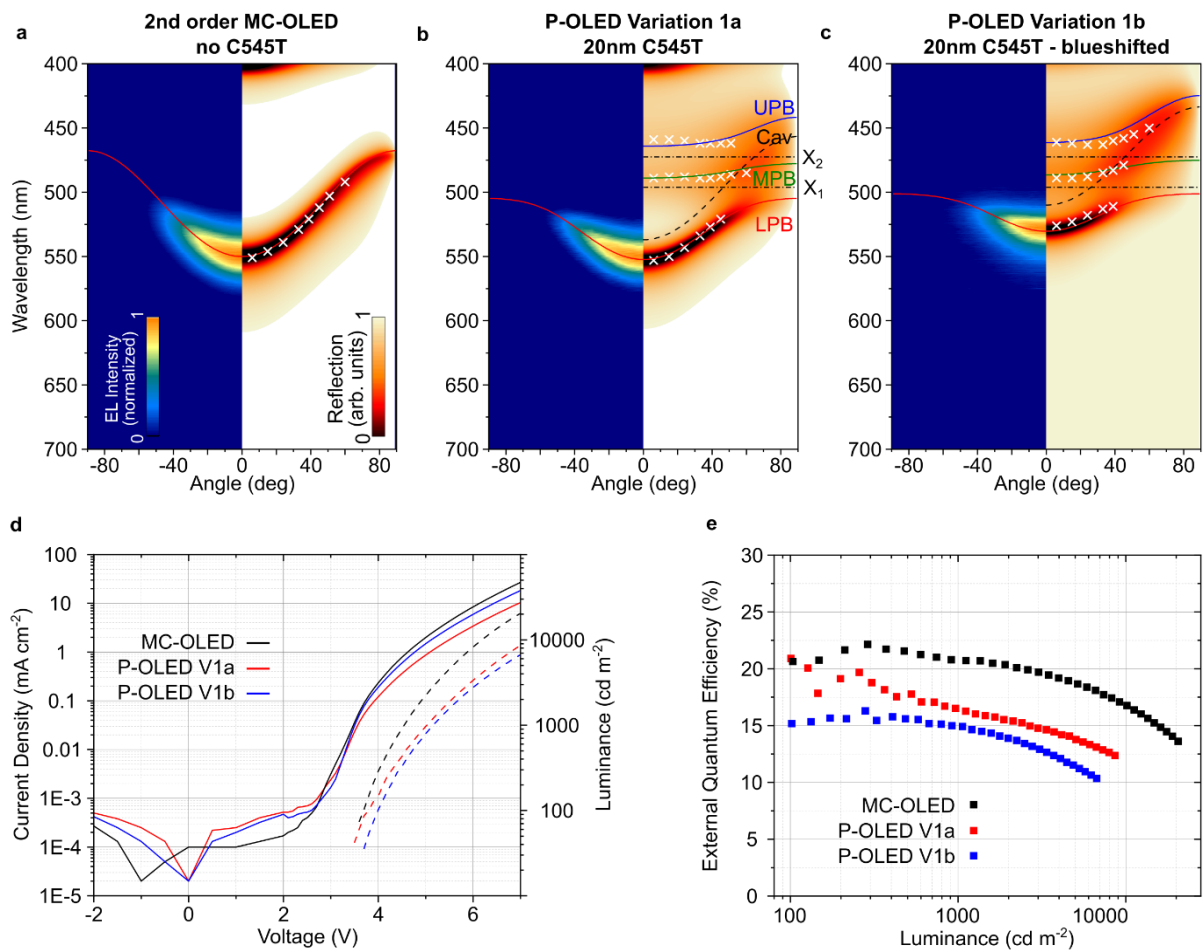

**Figure 2. Comparing microcavity OLED to P-OLED emission. a-c** Angle-resolved electroluminescence (left halves) and reflection (right halves) of a 2$^{nd}$ order MC OLED (**a**), a P-OLED with an SCL of 20 nm C545T (**b**), and a P-OLED with an SCL of 20 nm C545T and blue-shifted cavity (**c**). Both the emission as well as the reflection spectra clearly show the formation of polariton modes in **b** and **c**, including a reduced angular dispersion, in particular for the blue-shifted cavity. Polariton branches are modelled using a 3-coupled oscillator model (solid lines), coupling the original cavity resonance (dashed lines) and the two exciton resonances (dash-dotted lines). Crosses mark the local minima of measured angle-resolved reflection spectra. **d** Current density – luminance – voltage characteristics of the three designs. The introduction of the 20 nm SCL leads to a drop in current density (and luminance) due to the additional resistance introduced. **e** External quantum efficiency of the different OLEDs. The introduction of the SCL leads to a slight drop in efficiency for the non-optimized P-OLEDs.

To enhance light-matter interaction, we add a layer of C545T into the thick HTL, leading to the following configuration: 70 nm SpiroTTB:F6TCNNQ | 20 nm C545T | 40 nm SpiroTTB:F6TCNNQ (P-OLED Variation 1a). For this structure, we observe a splitting of the cavity resonance into a lower, middle and upper polariton branch in the measured and simulated reflection spectrum (Fig. 2b, false color – simulated, crosses - measured). Modelling of the polariton branches with the coupled oscillator model yields a Rabi-splitting of 0.2 eV between LP and MP, comparable to previous results using C545T[22]. The electroluminescence spectrum closely follows the lower polariton branch, indicating that emission from CzDBA is efficiently converted into polaritons. Compared to the bare MC, the P-OLED dispersion shows a flattening at larger angles, as the LP approaches the $X_1$ resonance.

This flattening of the dispersion can be further enhanced by shifting the cavity resonance towards smaller wavelengths through a moderate decrease in thickness of the transport layers of the OLED. Fig. 2c shows a thinner variation of the P-OLED (Variation 1b), with a blue-shifted peak emission at 528 nm, a FWHM of 25 nm (compared to 35 nm for the MC-OLED and 30 nm for P-OLED V1a) and significantly flattened LPB (10 nm blueshift at 45°, compared to 40 nm for the MC and 23 nm for P-OLED V1a). This kind of tuning of the position and shape of the LPB is a useful tool to strongly modify the emission characteristics of P-OLEDs. To optimize angular dispersion and emission spectrum, adjusting the interplay between thickness, Rabi-splitting, and absorption and emission spectra of the used dyes is crucial. For the combination of CzDBA and C545T used in the present study, we were able to optimize the system for emission at 528 nm, which is close to the ideal green color point in ultra-high-definition display applications (as given by recommendation BT.2020).

The performance of our MC- and P-OLEDs is summarized in Fig. 2 d,e. From the current density – voltage – luminance (*jVL*) characteristics, it is clear that the introduction of an undoped, bulk SCL on the hole side of the device adds bulk resistance and an energy barrier for hole injection, which, in combination, results in a decrease in current density and in turn a lower luminance at a fixed voltage. The EQE of the P-OLEDs drops by 20-30% in comparison to the MC-OLED. Nevertheless, P-OLED V1a still reaches an EQE of 16.5% at a luminance of 1,000 cd m$^{-2}$, while the blue-shifted V1b reaches an EQE of 15%. Both devices clearly surpass the best P-OLED performances reported previously[22].

**Optimizing the SCL design**

In order to further improve device performance, we investigate three different variations in the P-OLED design, in particular in the structure and implementation of the SCL. As shown for less efficient phosphorescent P-OLEDs before[22], an increase in coupling strength can aid in producing a flat dispersion while also achieving a higher device efficiency. By doubling the thickness of the C545T-based SCL to 40 nm, we roughly double the Rabi-splitting to 0.4 eV (due to a complex interplay between SCL thickness and electric field overlap this dependence deviates from the expected square-root proportionality). The structure and spectral characteristics of the resulting P-OLED (P-OLED V2) are presented in Fig. 3a, with the device performance shown in Fig. 3d,e. Despite a redshift in emission compared to P-OLED V1a, we observe a slightly reduced FWHM of 28 nm and a smaller blueshift at 45° of only 22 nm for this device, both of which are a result of the increased coupling strength. In addition, the CzDBA emission is more efficiently funnelled into the LPB at all angles, leading to an increased EQE of 21.3% at 1,000 cd m$^{-2}$, on par with the EQE of the MC-OLED. The insertion of a thicker bulk C545T layer, however, leads to a further decrease in current density (and thus forward luminance at a fixed voltage).

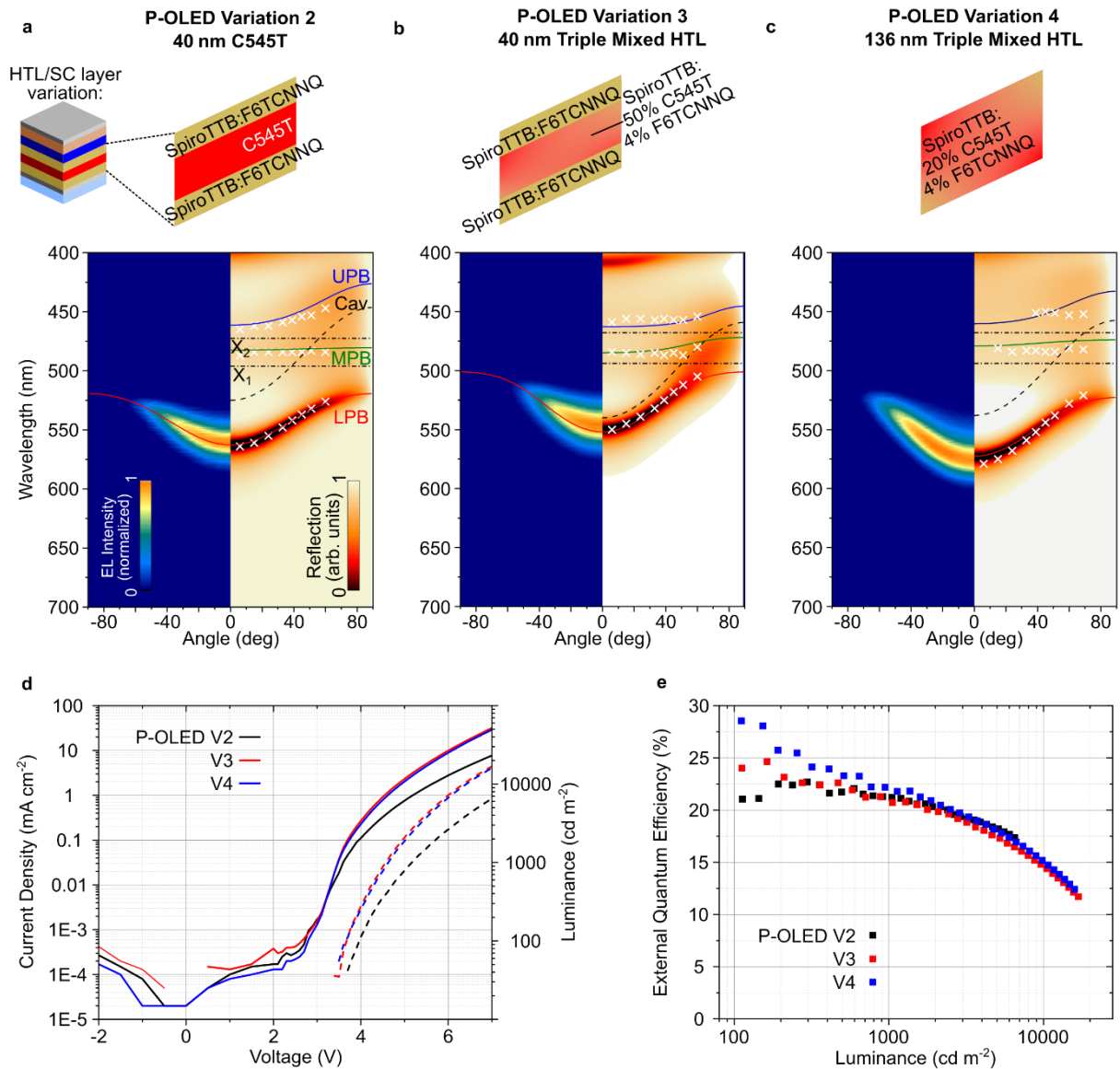

**Figure 3. Variations of TADF P-OLEDs. a-c** Design (top), and angle-resolved luminescence (left halves) and reflection spectra (right halves) of variations of the SCL design for TADF P-OLEDs, including modelled polariton branches (solid lines), cavity resonance (dashed lines) and exciton energies (dash-dotted lines). Crosses mark the local minima of measured angle-resolved reflection spectra. **a** Variation 2: increased thickness of a bulk C545T layer (40 nm) sandwiched between electrically doped HTLs. The increased SCL thickness leads to an increase in coupling strength and larger Rabi-splitting between the polariton branches. The resulting P-OLED shows a reduced angular dispersion and increased efficiency compared to P-OLED Variation 1a. **b** Variation 3: To avoid the additional resistance of the bulk C545T SCL, a mixed p-doped SCL consisting of SpiroTTB:50wt%C545T:4wt%F6TCNNQ is introduced, leading to the formation of polariton branches without noticeable loss in current density. **c** Variation 4: The doped HTL is completely replaced by a triple-mixed doped SCL consisting of SpiroTTB:20wt%C545T:4wt%F6TCNNQ (136 nm), facilitating both good conductivity and high coupling strength. This mixed layer leads to a redshift of the cavity mode even though the total cavity thicknesses was kept comparable to the other devices. **d,e** Current density – luminance – voltage characteristics (**d**) and external quantum efficiency (**e**) of the P-OLED variations. The optimized P-OLED variations achieve high efficiencies of >21% at 1000 cd m$^{-2}$, thus showing no loss compared to the weakly coupled MC OLEDs (Fig. 2).

To counteract the comparatively low hole conductivity of the C545T SCL, we explore the possibility of electrical doping. By thermal co-evaporation from three sources, we are able to create a mixed film of SpiroTTB:C545T:F6TCNNQ at precisely controlled mixing ratio. In this way, hole transport can be guaranteed through the high mobility of SpiroTTB[29] and the p-doping induced by F6TCNNQ, while the strength of light-matter interaction can be tuned through the mixing ratio of C545T. Importantly, C545T does not represent a significant hole trap as its HOMO (5.6 eV) is deeper than the HOMO of SpiroTTB (5.2 eV). Using the triple mixed HTL/SCL, we design further variations of our P-OLED. P-OLED V3 utilizes the SCL structure: 70 nm SpiroTTB:F6TCNNQ | 40 nm SpiroTTB:50wt% C545T:4wt% F6TCNNQ | 40 nm SpiroTTB:F6TCNNQ, with the resulting angle-resolved electroluminescence and reflection spectra shown in Fig. 3b. We observe clear formation of polariton branches, with a Rabi-splitting comparable to P-OLED V1a (20 nm neat C545T), LPB emission at 546 nm at 0°, an FWHM of 31 nm and a blueshift at 45° of 22 nm. Interestingly, the current density is strongly increased compared to the P-OLED variations with neat C545T (by a factor of 3-4), and now again comparable to the MC-OLED without SCL, while retaining a high EQE of 20.8% at 1,000 cd m$^{-2}$. This confirms the advantageous properties of a triple-mixed HTL/SCL combination, which shows excellent hole transport properties even when adding 50wt% of C545T.

In a final SCL variation, we exchange the entire HTL by a 136 nm-thick triple-mixed film of SpiroTTB:20wt% C545T:4wt% F6TCNNQ (V4, Fig. 3c). The increased total amount of C545T again leads to an increase in Rabi-splitting between LPB and MPB to 0.4 eV, comparable to P-OLED V2 (with 40 nm neat C545T). Due to the increased average refractive index and increased coupling strength, the LPB is redshifted, showing an emission peak of 573 nm at 0°, an FWHM of 33 nm, and a blueshift of 29 nm at 45°. The significant red detuning of the MC of this device leads to a considerable dispersion in the LPB, something that could be remedied by decreasing the thicknesses of the SCL or other transport layers. Despite the thick SCL, however, the P-OLED shows a high current density, comparable to the reference MC, and exhibits the highest EQE among the P-OLED variations tested here, with 22% at 1,000 cd m$^{-2}$, even reaching 28.6% at 100 cd m$^{-2}$.

The performance of all P-OLED variations is summarized in Table 1. The successful realization of all these different variations demonstrates the flexibility of the assistant SCL concept for realizing highly efficient polariton emission, without showing major drawbacks over reference devices based on the same EML and similar device architecture. The results demonstrate that careful design and tuning allow to obtain P-OLEDs with high efficiency, narrowband emission and low angular dispersion.

**Top-emitting polariton OLEDs**

While bottom-emitting OLED designs are well studied and often simpler to fabricate, top-emitting designs (i.e. showing light emission through the cathode) have significant advantages for application in displays. A top-emitting design offers facile integration with non-transparent substrates, such as complementary-metal-oxide-semiconductor (CMOS) driver backplanes for high-resolution displays and improves fill-factor when used with the thin-film transistor (TFT) backplanes of more conventional displays. The P-OLED architecture is ideal for this configuration, as it does not include transparent conductive oxide electrodes, which generally require deposition under conditions that can be harsh to an underlying organic film. In order to realize a top-emitting-design P-OLED, we utilize a thick (100 nm) Ag anode and a thin cathode comprising 1 nm LiF | 1 nm Al | 20 nm Ag layers, as well as a 60 nm NPB capping layer, to aid outcoupling of light at the Ag-air interface. Through transfer-matrix and outcoupling simulations, all thicknesses are optimized to achieve high efficiency, narrowband emission and low angular dispersion simultaneously.

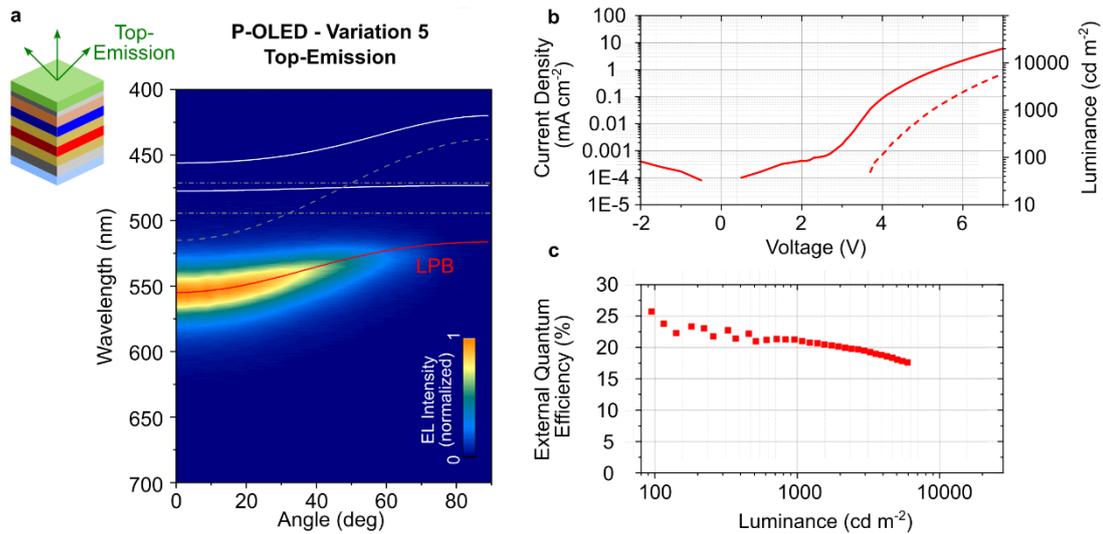

**Figure 4. Top-emitting P-OLED. a** Top-emitting structure and angle-resolved electroluminescence of the TADF P-OLED. To facilitate top-emission, the previously used cathode is replaced by a 1 nm LiF | 1 nm Al | 20 nm Ag electrode design, and a capping layer of 60 nm NPB is added to aid light outcoupling at the cathode-air interface. The SCL consists of 40 nm C545T sandwiched by SpiroTTB:F6TCNNQ HTLs (compare Variation 2). All layer thicknesses are optimized for efficient top-emission and flat dispersion along the lower polariton branch (solid red line). **b,c** Current density – luminance – voltage characteristics (**b**) and external quantum efficiency (**c**) of the top-emitting P-OLED.

The structure and characteristics of the optimized top-emitting P-OLED are shown in Figure 4. The emission follows the lower polariton branch, showing a peak at 555 nm at 0°, an FWHM of 31 nm and a blueshift of 19 nm at 45°. The top-P-OLED exhibits an EQE of 21.2% at 1,000 cd m$^{-2}$ and reaches 25% at 100 cd m$^{-2}$. As demonstrated for P-OLED variations 1a and 1b, the angular dispersion can be further reduced by blue-shifting the cavity. For example, a top-P-OLED with a peak emission shifted to 541 nm exhibits an angular blueshift of less than 10 nm at 45° and an FWHM of 27 nm.

The top-P-OLED design could be further enhanced by utilizing the triple-mixed SCL and a further increase in coupling strength by increasing the amount of C545T used. We expect that this will provide ideal conditions for the green pixels in ultra-high-definition OLED displays.

**Table 1**. Summary of device performance for main OLED variations.

|  | λ (nm) | FWHM (nm) | Δλ at 45° (nm) | EQE at 1,000 cd m$^{-2}$ (%) | Current density at 6 V (mA cm$^{-2}$) |
|---|---|---|---|---|---|
| **MC-OLED** | 550 | 35 | 40 | 20.8 | 8.4 |
| **P-OLED V1a** | 550 | 30 | 23 | 16.5 | 3.4 |
| **P-OLED V1b** | 528 | 25 | 10 | 15.0 | 5.9 |
| **P-OLED V2** | 560 | 28 | 22 | 21.3 | 2.8 |
| **P-OLED V3** | 546 | 31 | 22 | 20.8 | 9.9 |
| **P-OLED V4** | 573 | 33 | 29 | 22.0 | 8.9 |
| **P-OLED V5** | 555 | 31 | 19 | 21.2 | 2.2 |

## Discussion

MC OLEDs are of great interest to the display sector due to their facile fabrication, superior mechanical flexibility compared to designs that rely on transparent conductive oxides, and compatibility with various backplane technologies, such as high fill-factor TFT and CMOS, the latter being particularly relevant for OLED microdisplays[30]. While traditional designs require minimizing the reflectivity of the light-emitting electrode, as highly reflective electrodes lead to undesirable color shifts when viewed from different angles[31], P-OLEDs offer an arguable more attractive solution by enabling the use of highly reflective electrodes while also enhancing color consistency across viewing angles and improving color saturation. These more reflective electrodes also bring practical advantages, such as simplifying the fabrication process by reducing the likelihood of metal nano-island formation and boosting electrical conductivity.

We have demonstrated the successful combination of an extremely efficient yet intrinsically broadband TADF emitter with a strong coupling architecture through the use of a strongly absorbing assistant SCL embedded within the HTL. By boosting the strength of light-matter interaction within the cavity, the devices entered the strong coupling regime and showed exclusive emission from the lower polariton branch, suggesting a highly efficient conversion of CzDBA emission into polaritons. The assistant SCL design was made possible by incorporating a doped hole transport layer, which significantly increased the cavity thickness and enabled the structure to reach the total thickness required for a second-order resonance. By separating the EML and SCL and placing each in an electric field maximum of the cavity mode, we maximized both coupling strength and external quantum efficiency. The resulting devices showed emission spectra that are strongly narrowed compared to the bare material emission, but also compared to conventional MC OLEDs in the weak coupling regime that use the same emitter[32]. Optimized designs achieved external quantum efficiencies of above 20% (at 1,000 cd m$^{-2}$), thus at least doubling the efficiency over the previous record for P-OLEDs[22]. At a lower luminance of 100 cd m$^{-2}$, our P-OLEDs even reached efficiencies up to 28%, becoming competitive to conventional non-cavity designs with broadband emission. By fabricating and testing different variations in our SCL architecture and in particular through the use of a triple-mixed SCL/HTL, we were able to overcome the challenges previously faced by P-OLEDs, and now reach similar efficiencies, current densities and luminance as conventional MC designs, but with a narrowed linewidth and a drastically reduced angular dispersion.

Considerable research effort is dedicated to developing new materials with intrinsically narrow emission linewidths. Polaritonic cavities offer an alternative, and potentially complimentary strategy to achieving narrowband emission, independent of emitter material. The assistant SCL architecture offers tuneability to optimize a wide range of different emitters for use in displays with high color brilliance, without losing efficiency or brightness. In the future, emission linewidth might be further reduced by increasing the cavity strength through an increase in electrode reflectivity, e.g. by increasing the thickness of the semi-transparent electrode, or adding distributed Bragg reflectors[33,34].

Our current work demonstrates, for the first time, that the polariton OLED approach can be effectively applied to TADF emitters and can surpass the critical EQE threshold of 20%. Given that this architecture closely resembles existing OLED structures and only requires the addition of a single new material compared to standard TADF designs, we believe its adoption in commercial OLED displays could be achieved with relatively minor effort.

## Methods

*Sample fabrication:* OLEDs were fabricated via thermal evaporation of organic and metal thin films at a base pressure of $1 \times 10^{-7}$ mbar (Angstrom EvoVac) onto 1.1 mm-thick glass substrates. The materials used were Al and Ag as metal electrodes (Kurt J. Lesker Co.), MoO$_3$ and LiF as injection layers, 2,2',7,7'-tetrakis(N,N'-di-p-methylphenylamino)-9,9'-spirobifluorene (Spiro-TTB), 1,3,4,5,7,8-hexafluorotetracyanonaphthoquinodimethane (F6TCNNQ), 2,3,6,7-tetrahydro-1,1,7,7,-tetramethyl-1H,5H,11H-10-(2-benzothiazolyl)-

quinolizino-[9,9a,1gh]-coumarin (C545T), N,N′-di(naphtalene-1-yl)-N,N′-diphenylbenzidine (NPB), tris(4-carbazoyl-9-ylphenyl)amine (TCTA), 4,4'-Bis(carbazol-9-yl)biphenyl (CBP), 9,10-bis(4-(9H-carbazol-9-yl)-2,6-dimethylphenyl)-9,10-diboraanthracene (CzDBA), and 1,3,5-Tri[(3-pyridyl)-phen-3-yl]benzene (TmPyPB). All organic materials, LiF and $MoO_3$ were obtained from Lumtec in sublimed grade and used as received. Film thicknesses were controlled in situ using calibrated quartz crystal microbalances (QCMs). Before fabrication, substrates were cleaned by ultrasonication in acetone, isopropyl alcohol and deionized water (10 min each), followed by UV-ozone cleaning for 3 min. OLEDs were encapsulated in a nitrogen atmosphere with a glass lid using UV-curable epoxy (Norland NOA68), preventing intermittent exposure to ambient air. The active area of these devices was 4.0 $mm^2$.

*Device characterization:* Device structures and electric field distributions of the OLEDs were simulated using a transfer matrix model. Polariton branches were calculated using a coupled oscillator model[11]. Current density and luminance–voltage behaviour were analysed with a source measure unit (Keithley 2450) and a calibrated amplified Si photodiode (Thorlabs PDA100A) positioned at a fixed distance of 16.7 cm and connected to a digital multimeter (Keithley 2100). Angle-resolved spectra were measured using a goniometer setup built in-house[35] equipped with a fibre-coupled spectrometer (Ocean Optics HDX VIS-NIR). The EQE was calculated by taking into account the angular emission characteristics of the OLED[36]. For the top-emitting design, the angle-resolved transmission of the encapsulation glass lid was measured to correct the OLED spectral radiant intensity for the Fresnel-losses occurring in the glass. Angle-resolved reflectivity spectra were recorded using a UV-VIS-NIR spectrometer with an automated rotating sample and detector unit (Agilent Cary 6000i with Universal Measurement Accessory).

## Acknowledgements

This research was financially supported by the Alexander von Humboldt Foundation (Humboldt Professorship to M.C.G.), the European Research Council under the European Union's Horizon Europe Framework Program/ERC Advanced Grant agreement no. 101097878 (HyAngle, to M.C.G.) and the Deutsche Forschungsgemeinschaft (Research Training Group "TIDE", RTG2591). A.M. acknowledges funding from the European Union's Horizon 2020 research and innovation program under the Marie Skłodowska-Curie grant agreement no. 101023743 (PolDev, to A.M.) and from the Bundesministerium für Bildung und Forschung (BMBF) within a GO-BIO initial project no. 16LW0454 (FluoPolar, to A.M.).

## Author Contributions

A.M. and M.C.G designed the study. S.L. and A.M. fabricated and characterized all POLED devices, with help from J.W. and S.H.. V.G. and F.T.C. optimized reference and microcavity OLEDs. A.M. performed transfer matrix and coupled oscillator modelling. All authors evaluated and discussed the results. A.M. and M.C.G. wrote the manuscript with input from all authors.

## Competing Interests

A.M. and M.C.G. are inventors of a pending patent application filed by the University of Cologne on TADF P-OLEDs. The remaining authors declare no competing interests.